# A Generic Data Acquisition Framework For High Performance 2D X-RAY Detectors

Wassim Mansour, *Member, IEEE,* Nicolas Janvier, *Member, IEEE,* and Pablo Fajardo

*Abstract*—This paper presents the design criteria and the current implementation of a generic and functionally rich data acquisition framework for high performance detectors called RASHPA. The framework is based on the use of RDMA mechanisms for optimized data transfer and supports multiple destinations and simultaneous transfer operations through multiple parallel data links. Although RASHPA is somehow agnostic in what respects to the type of detector and can deal with different types of data and metadata, it implements selection and dispatching rules that are optimized for the efficient manipulation and distribution of images. For all the previous reasons, the full potential of RASHPA comes up when implemented in very high data throughput modular 2D detectors as most of the advanced new area detectors that are in development for synchrotron and free-electron laser applications.

## I. Introduction

DATA production and analysis are the essential components at the heart of any scientific experimental application. The produced data rates by x-ray detectors increases significantly, as the technology of the photon sources evolves [1].

Modern high performance 2D detectors for photon science applications are able to produce very high throughput in the range of 1 to 100GBps. The Eiger 9M [2] and Jungfrau 10M [3] detectors developed at the Paul Scherrer Institute (PSI) with throughputs that go up to 360 Gbps and 400 Gbps respectively are good examples of this type of advanced detectors. Transferring, manipulating and processing such data streams in acceptable times is a severe challenge.

Traditionally, efforts on detector development for photon sources have focused on the properties and performance of the detection front-ends. In many cases, the data acquisition chain has been treated as a complementary component of the detector system that was added at a late stage of the project. In some cases, the data acquisition subsystems, although achieving minimum bandwidth requirements were kept relatively simple in term of functionalities, in order to minimize design effort, complexity and implementation cost.

This approach is changing in the last years as it does not fit the new high performance detectors; industrial data acquisition protocols do not provide the required data throughput and implementing high performance schemes becomes much more difficult and resource consuming. Detector developers are changing their paradigm and moving into the development and implementation of reusable high performance data acquisition schemes that can be applied to different kind of detector devices [4].

The design and implementation of efficient data acquisition scheme with multi-gigabyte per second capabilities become a mandatory and unavoidable solution to deal with such data rates. RASHPA (RDMA-based Acquisition System for High Performance Applications) is the generic data acquisition framework currently under development at the ESRF. It is optimized for the transfer of 2D detector data, i.e. images, metadata, etc., relying completely on Remote Direct Memory Access (RDMA) mechanism. Therefore, RASHPA is able to push data, at maximum throughput into the address space of one or several destination backend computers. RASHPA scheme provides a high standardization level in the data transmission pipeline from the detector up to the software application for further processing, visualization or storage.

This paper details RASHPA conceptual design and is organized as follows: section II states the objectives of the data acquisition platform. Section III, details the concept behind RASHPA. In section IV, the hardware resources requirements in order to implement a RASHPA compatible system are discussed. The overall system functional description is presented in section V. Section VI discusses the software library used to configure the platform. Section VII shows results of the first two prototypes and section VIII concludes and provides some perspectives.

## II. Objectives

The design of RASHPA framework pursues three main objectives discussed in the following subsections.

### A. Promoting standardization and reusability

The design and implementation of an efficient data acquisition schemes with multi-gigabyte per second

Paper submitted for review on put date here
This project has received funding from the European Union's Horizon 2020 research and innovation programme under grant agreement No. 654220.
W. Mansour is with the European Synchrotron Radiation Facility, Grenoble, FRANCE; mail: wassim.mansour@esrf.fr
N. Janvier is with the European Synchrotron Radiation Facility, Grenoble, FRANCE; mail: nicolas.janvier@esrf.fr
P. Fajardo is with the European Synchrotron Radiation Facility, Grenoble, FRANCE; mail: pablo.fajardo@esrf.fr

capabilities and high-level functionality is far from trivial. Therefore, it is of overriding importance that the invested effort can target a variety of detector systems and over a long-lasting period of several decades. One of the key reasons behind that is the long-term development cycles for advanced X-ray detectors that can easily take a decade from concept to final instrument.

With this purpose, the data acquisition scheme must focus on conceptual and functional aspects and should minimize the dependency on a given technology. This is particularly important in what respects data link hardware and protocols so that RASHPA should be able to adopt future standards, following and profiting from the evolution of data communication technology.

The proposed framework should be independent of the particularities of the detection front-end and other instrument features that are not directly related to the data transmission process. In this way it must be possible to design detectors, in which, the RASPHA related functionalities could be included in a reusable functional block with well defined interfaces.

In order to simplify the system development without compromising the performance target, the rich functionalities provided by RASHPA must be managed at the backend side by detector independent code that provides an interface sufficiently high-level to be easily exploited by the diversity of detector software applications that may be used in the full detector systems. This approach is expected to greatly reduce software development efforts and make the data acquisition framework much more attractive for detector developers.

*B. Addressing a wide range of implementations*

Most detector developments for photon sources do not target a single scientific application. It is therefore frequent and convenient that the same detector technology and components are used for various applications. These applications may have rather different data-throughput and performance requirements as well as their own set of technical and cost constraints. In practice, this is addressed by producing different implementations of the detection systems. It is therefore very important that a generic data acquisition framework as RASHPA can fit in such a scheme and that the future compliant detectors are not restricted only to very end configurations.

Scalability is therefore a major issue. In the most demanding experiments, it must be possible to build high performance RASHPA based detectors able to send data to computing farms by using large number of the fastest data links available at the time of the development. On the other hand, it must be possible to scale down the systems to rather simple and relatively inexpensive configurations. One important goal behind RASHPA is that the system could be rescaled by reconfiguration of the components and not requiring any hardware redevelopment.

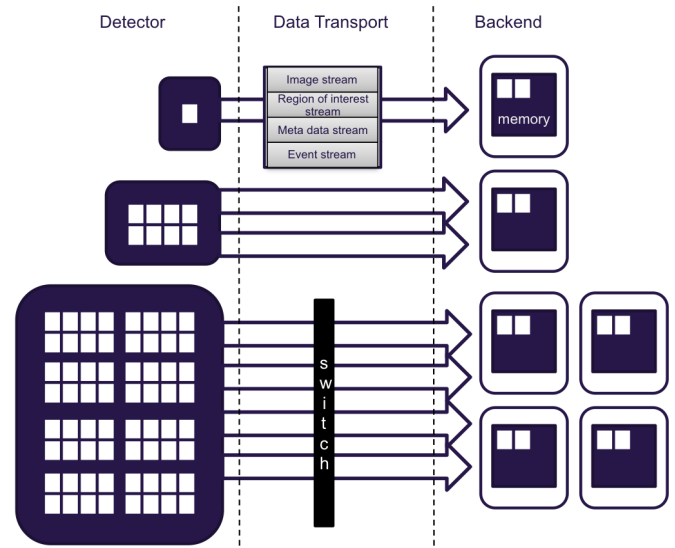

Fig. 1. Block diagram of Rashpa's architecture

*C. Alleviating the data management challenge*

Moving data at high rates from the detector head to memory buffers in the computer infrastructure is a necessary step to take advantage of many of the most advanced detector systems. This does not solve other difficulties related to data management and manipulation. The effective exploitation of such very high-throughput data streams is extremely challenging and must consider all the on-line data management aspects from data generation and transmission, to data dispatching, visualization, analysis and storage.

Therefore, RASHPA data acquisition framework should support the following features depicted in figure 1:

1) Multiple sources and destinations via routable network topology based on high speed data links.

2) Parallel and simultaneous data streams.

3) Remote direct memory access (RDMA) zero copy transfer features

4) Potential support of image manipulations and other hardware accelerator algorithms at hardware level.

As previously mentioned, RASHPA is conceived to serve as basis in the design of the data acquisition mechanism for a diversity of detectors. For the sake of reusability and longevity, the framework itself imposes very few constraints on the choice of technologies employed for the implementation of the actual detector devices. With this approach, it is expected that RASHPA will be applicable to a larger scope of detector developments and will take advantage of future progress in electronics and data communication technology. It is also foreseen that detector software implementations that use RASHPA as its internal machinery for data acquisition purposes will provide interoperability with different detector designs, and will be usable with various generation of detectors.

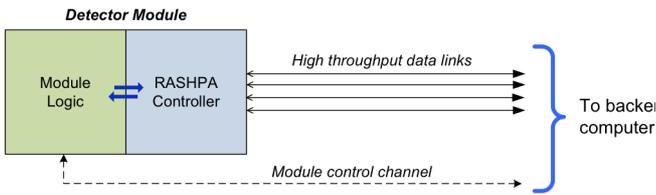

Fig. 2. Basic scheme of Rashpa detector module

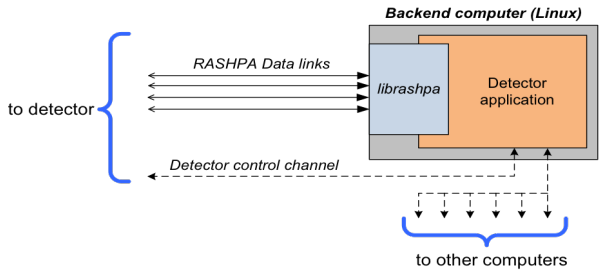

Fig. 3. Basic scheme of Rashpa backend computer

### III. Key Concept

Any implementation of RASHPA relies on the following main components:

1) Detector Module (DM): is the basic element building the detectors that includes a functional block, called RASHPA Controller (RC) responsible of transferring data to the Backend Computers (BC) via high-speed data links. Figure 2 shows a basic scheme of a RASHPA detector module.

2) Multi Module Detector (MMD): is the complete detector consisting of one or more identical DMs.

3) Backend Computers (BC): These are the final destinations of the data. In RASHPA terminology, all the computers that receive detector data are called data receivers (DR). The backend computer that is in charge of the configuration and initialization of the data acquisition subsystem is called the system manager (SM). Figure 3 shows a basic scheme of a RASHPA backend computer. Each backend computer runs software integrating a C-language library called LIBRASHPA, that is responsible of managing and implementing the specific RASHPA functionalities.

4) RASHPA Telegrams (RT): XML telegrams used to retrieve the capabilities and configuration of the DMs and DRs. The RTs are generated and distributed to the concerned components during the configuration phase.

5) RASHPA Address Space (RAS): it is a single common address space that remaps all the memory areas of all the DRs. It appears to be a contiguous block even if it may include regions from more than one backend computer. It could also combine system RAM memory buffers with other memory areas such a memory in coprocessor boards (e.g. GPUs). It is constructed by dedicated functions in the RASHPA software library.

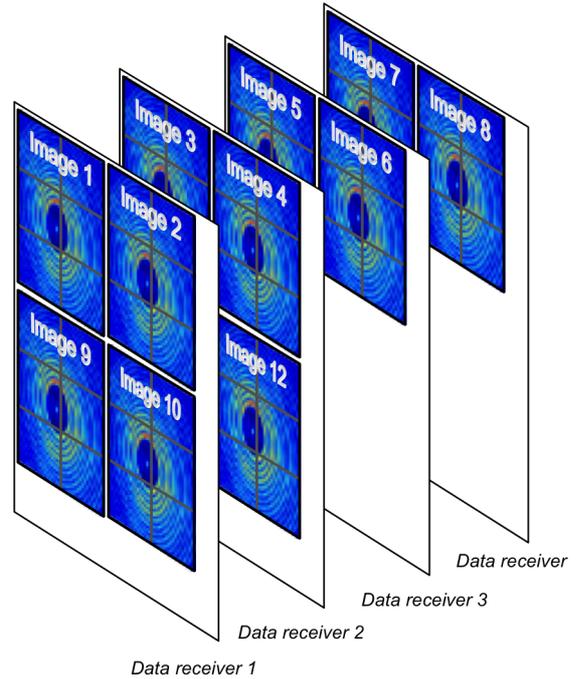

Fig. 4. Example of a DTP result from a segmented 2D detector composed by six independent modules to four RASHPA BCs

6) RASHPA Buffers (RB): consists of a set of address areas or blocks named local buffers (LB) that reside in the address space of one or more DR. The simplest case of RB would be a conventional buffer declared in the system RAM area of one DR.

7) Data slices, data sets and data blocks: Detector data are organized in sequences, produced and numbered consecutively, called data slices (DS). A DS can be consisted of data produced by DM during a time interval.

Every type or class of data produced by DM and treated separately is called a dataset (ex. image, metadata, etc.).

All the data from a given dataset, produced within a data slice is called a Data Block (DB).

8) Data Transfer process (DTP): is a description of the direct transfer from DM to RB in a given DR. It supports RDMA and guarantees data integrity. In other words, DTP specifies what to send, where to send and how to send the data. An example of the result of a DTP from a segmented 2D detector composed by six independent modules to four RASHPA BCs is illustrated in figure 4.

9) Data Channels (DC): These are functional RASHPA blocks, responsible for data transfer from the DMs to the destination buffers. A DC is configured as a part of a DTP. In most general case, a DTP requires the activation of at least one DC for each DM.

10) Event channels: These are functional RASHPA blocks responsible of sending events to the BCs. Events are

asynchronous messages generated by the RC in the DM. They are used to signal the detector software application about errors, change of status, progress of the running DTP, etc.

## IV. HARDWARE RESOURCE REQUIREMENTS

In order to build a RASHPA compliant detector and software application, some requirements should be respects at the data link, the detector module and the backend computers.

### A. Data Links

The transfer of data between the detector modules and the backend computers is achieved by fast links that fulfill the following requirements:

1) Direct memory access (DMA): The full address space of the backend computers should be accessible from the RASHPA controllers by RDMA mechanisms. In RASHPA scheme, the effort and complexity is put on the initialization and configuration of the system; once the data transfer operations started there is no need of any intervention by the CPUs in the data receiver.

2) Asynchronous event signalling: The selected data links must provide a low-level asynchronous mechanism to signal conditions that trigger RASHPA events. The conditions activated in each case are selected by LIBRASHPA as part of the DTP configuration.

3) Data integrity: RASHPA requires that the data link layer implement all the mechanisms necessary to achieve data integrity during the transfer process. It may include any packet ordering, retransmission or error correction schemes as needed.

4) Bidirectional operation: The data links operate as unidirectional channels from the detector towards the backend computers. However, the use of bidirectional links would allow future functional extensions of the framework such as high-level flow control mechanism or embedded control channels. In any case, the data bandwidth requirements would be very asymmetric and in order to minimize misuse of the links, they should be used as write only channels and avoid read operations that may slow down or increase the latency of the system.

5) Data switches: The link technology must be compatible with the implementation of data switches for multiple-host configurations. The data switches must provide mechanisms to both write data and transmit asynchronous events to the backend computers.

There are no specific requirements on data bandwidth or other minimum performance figures for the capacity of the data links. However, as the main goal of the RASHPA framework is to allow very high throughput data transfers. It is expected that the RASHPA based systems will use advanced data links for which the implementations will evolve in the future following the evolution of the data communication technology.

From a purely functional point of view, the link technology considered the most suitable to be used in a RASHPA system is PCI Express over cable [5], and it was the choice adopted for the implementations of the first prototype. PCIe is the natural extension of the internal bus of the backend computers and it does not need any software protocol. The hardware and the operating systems of the computers take in charge the full initialization of the data links that provide data integrity and minimum latency. In addition, it is possible to integrate PCIe endpoints in the detector modules with a reasonable design effort. Disadvantages of using this link includes the limited availability of commercial off-the-shelf data switches and PCIe links based on optical cables as well as the small packet transfer size. Therefore, other candidates based on more widespread technologies, such as Infiniband [6] or Ethernet, are under consideration athough Ethernet, for instance, needs to be extended with remote DMA protocols (RDMA) such as iWARP[7] or RoCE [8-9].

### B. Detector Modules

The RASHPA framework specifies the functionality of the detector modules but not the internal resources required to achieve such functionality.

In the case of DM with multiple data links, the DM must either, know in advance or detect at runtime, how many data links are physically connected and in operational state. A module must be able to operate even if only part of the data links is operational during DTP initialization. All the operational data links must be able to access the full RAS. The framework does not specify how the data write operations are shared among the various links, this is the responsibility of the module developers.

A detector module must implement the required functionality of a RASHPA controller and integrate, in addition to the fast data link interfaces, high-level configuration features and powerful data manipulation capabilities.

### C. Backend Computers

All the backend computers in a RASHPA based data acquisition system must be based on the Linux operating system and run detector software that include and use one instance of the LIBRASHPA library. The backend computer that acts as SM plays a central role for initialization, configuration and overall system monitoring. The SM must manage or at least have access to the control link.

All the BCs must also include the data link hardware interfaces that are required to implement the configuration selected for the particular application. Both the SM and DR must be able to accept and treat the asynchronous messages from the DMs that trigger RASHPA events.

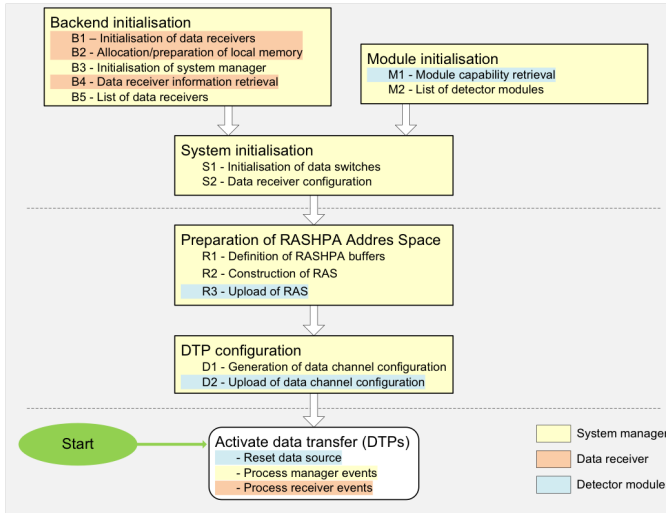

Fig. 5. Functional steps in RASHPA system

The DRs must be able to map the RASHPA buffer in their system address space. The data destination buffers will be in most of the cases large memory areas in system RAM but they may also consist of buffers in extension boards such as GPU or FPGA coprocessors or disk controllers.

## V. System Functional Description

The RASHPA framework defines the conventions, procedures and functionality that a compliant data acquisition system must implement. DMs must be configured at the initialization phase and operate as expected by the DTP. At the backend side, the detector application software uses the high-level specific functionalities provided by LIBRASHPA to comply with the framework. It is the responsibility of the detector applications to orchestrate the full operation of the complete system. This is done by properly combining the control of the detector with calls to LIBRASHPA functions. The detector application is also responsible of the intercommunication with various computers in the case of configurations with multiple backends.

Figure 5 illustrates the different steps and partial operations in a RASHPA system. The figure shows the dependencies between those steps and the order in which the detector application has to handle them.

When a detector application repeats acquisition and data transfer sequences, not all the operations have to be done again and the application will normally loop across the operation flow in figure 5.

The initialization of the system must start by the initialisation of the main system components: the backend computers (B1 to B5) and all the detector modules (M1, M2). Once all those individual components are initialised the system manager may proceed to complete the this first phase by initialising the data switching system (S1) and dispatching the configuration of the communication system to all the data receivers (S2). At this point the system is ready to start the configuration phase.

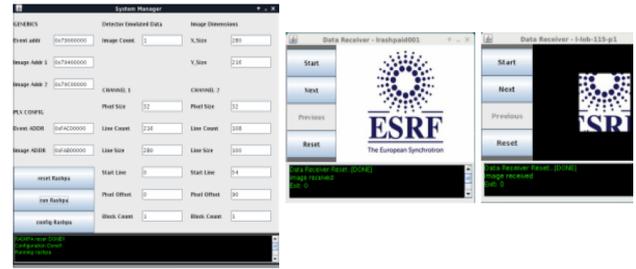

Fig. 6. Results for a PCIe over cable RASHPA network

The first step in the configuration phase if the definition of the RASHPA buffers (R1) that will be used to receive the data streams during the transfer operations. This step is necessary to generate and upload the RAS map (R2, R3) into all the detector modules. The next configuration step is the generation of the configuration of the DTPs (D1) that must be uploaded into the detector modules (D2).

At this point the system is ready to start data transfer by activating the configured data channels.

## VI. Software Library (LIBRASHPA)

LIBRASHPA is the piece of software in charge of managing all the specific aspects of a RASHPA based system. LIBRASHPA is a Linux library written in C language that must be included in the detector software applications such as the detector servers based the LIMA data acquisition and control library used at the ESRF and other synchrotron radiation facilities [10].

The main duties of a LIBRASHPA manager are:

1) Compiling the capabilities of all the DMs and the information from all the DRs in the system.

2) Initializing backend components, such as data switches in case of routable network.

3) Building and maintaining the RASHPA global address space.

4) Managing the configuration of the data reception buffers and provide the information required by the DRs.

5) Providing the configuration of the detector modules from the definition of the requested DTPs.

6) Managing and triggering system wide events and passing information to the detector application.

The duties of a LIBRASHPA for data receivers include allocating physical memory for data reception, triggering receiver events and passing the information to the detector application.

## VII. Prototype Results

The very first RASHPA demonstrator was developed in the frame of the European project CRISP [1]. In that first version, RASHPA supported the data transfer from a single DM to a single BC via PCIe over cable. It was tested and validated using a data generator emulating the detector behaviour.

In an advanced RASHPA prototype [11], a more complex RASHPA network has been constructed. It is also based on PCIe over cable. In order to build such network one would need to select proper PCIe switches, therefore, the PXH810 from DOLPHININCS was selected. It is based on PLX PEX8749 PCIe switch [12]. A complete reconfiguration of the plx chip was needed to make it RASHPA compatible. The full RASHPA implementation was tested using a data generator emulating the detector and storing the ESRF logo as a test image. In this implementation, only two data channels were activated, the first one configured to send the full image to one DR, whereas the second DC sends a region of interest of the same image to the second DR. Figure 6, shows the results obtained when performing the tests.

In addition to demonstrators and testing prototypes, RASHPA is being integrated in the SMARTPIX detector [13], a Medipix3 [14] based x-ray detector currently under development at the ESRF. The prototype validates the RASHPA concept when supporting a PCIe-over-cable routable network. Currently, tests and the FPGA implementation of other types of data links, such as RoCEv2 and an in-house developed RDMA over Ethernet protocol are under investigations.

## VIII. Conclusions and Future Work

The ESRF RDMA-based data acquisition platform, called RASHPA, still under development is presented in this paper along with a detailed introduction to each of the main components of the platform and their functional role.

The paper also presented the first two implementations of the RASHPA hardware, in order to prove the feasibility of the proposed platform. The first one was very basic data acquisition supporting a single detector module and a single backend computer whereas the second one is much more complex when dealing with routable PCIe over cable network.

Among the short term perspectives of the project there is the completion of LIBRASHPA, the library that implements and supports the software functionalities in RASHPA, and the application of the whole data acquisition framework to the new generations of detectors currently under development at ESRF.

In addition to that, it is also under investigation the integration of image manipulation algorithms within the RASHPA hardware in order to accelerate processes that consume substantial CPU resources, and the convenience of developing adaptor modules that may convert the data streams produced by some non-ESRF detectors and following other data transfer protocols into fully RASHPA compatible data acquisition schemes.